\documentclass[%
twocolumn,
superscriptaddress,
 amsmath,amssymb,
 aps,prx,
]{revtex4-2}

\usepackage{bbold}
\usepackage{amsfonts}
\usepackage{dsfont}
\usepackage{xcolor}
\usepackage{bm}
\usepackage{hyperref}
\hypersetup{plainpages=false,colorlinks=true,linkcolor=blue, citecolor=blue, urlcolor=blue}

\usepackage{xcolor}
\usepackage{bbold}
\usepackage{mathrsfs}
\usepackage{natbib}
\usepackage{here}
\usepackage{comment}

\usepackage{graphicx}

\newcommand{\tr}[1]{\textcolor{red}{#1}}
\newcommand{\tg}[1]{\textcolor{green}{#1}}
 
\usepackage{array,mathtools,amssymb,booktabs}
\newcolumntype{C}{>{$}c<{$}}

\begin{document}
\title{Hidden Markov model analysis to fluorescence blinking of fluorescently labeled DNA} 
\author{Tatsuhiro Furuta}
\affiliation{Graduate School of Engineering, Kyushu Institute of Technology, 1-1 Sensui-cho, Tobata-ku, Kitakyushu, Fukuoka 804-8550, Japan.} 

\author{Shuya Fan}
\affiliation{SANKEN (The Institute of Scientific and Industrial Research), Osaka University, Mihogaoka 8-1, Ibaraki, Osaka 567-0047, Japan.}

\author{Tadao Takada}
\affiliation{Department of Applied Chemistry, Graduate School of Engineering, University of Hyogo, 2167 Shosha, Himeji, Hyogo 671-2280, Japan.} 

\author{Yohei Kondo}
\affiliation{Department of Life Science and Technology, Institute of Science Tokyo, Nagatsuta, Midori-ku, Yokohama, Kanagawa 226-8501, Japan.} 

\author{Mamoru Fujitsuka}
\affiliation{SANKEN (The Institute of Scientific and Industrial Research), Osaka University, Mihogaoka 8-1, Ibaraki, Osaka 567-0047, Japan.} 

\author{Atsushi Maruyama}
\affiliation{Department of Life Science and Technology, Institute of Science Tokyo, Nagatsuta, Midori-ku, Yokohama, Kanagawa 226-8501, Japan.} 

\author{Kiyohiko Kawai}\email{kawai.k@bio.titech.ac.jp}
\affiliation{Department of Life Science and Technology, Institute of Science Tokyo, Nagatsuta, Midori-ku, Yokohama, Kanagawa 226-8501, Japan.} 

\author{Kazuma Nakamura}\email{kazuma@mns.kyutech.ac.jp}
\affiliation{Graduate School of Engineering, Kyushu Institute of Technology, 1-1 Sensui-cho, Tobata-ku, Kitakyushu, Fukuoka 804-8550, Japan.} 
\affiliation{Integrated Research Center for Energy and Environment Advanced Technology, Kyushu Institute of Technology, 1-1 Sensui-cho, Tobata-ku, Kitakyushu, Fukuoka 804-8550, Japan.}

\begin{abstract}
We examine quantitatively the transition process from emitting to not-emitting states of fluorescent molecules with a machine learning technique. In a fluorescently labeled DNA, the fluorescence occurs continuously under irradiation, but it often transfers to the not-emitting state corresponding to a charge-separated state. The trajectory of the fluorescence consists of repetitions of light-emitting (ON) and not-emitting (OFF) states, called blinking, and it contains a very large amount of noise due to the several reasons, so in principle, it is difficult to distinguish the ON and OFF states quantitatively. The fluorescence trajectory is a typical stochastic process, and therefore requires advanced time-series data analysis. In the present study, we analyze the fluorescence trajectories using a hidden Markov model, and calculate the probability density of the ON and OFF duration. From the analysis, we found that the ON-duration probability density can be well described by an exponential function, and the OFF-duration probability density can be well described by a log-normal function, which are verified in terms of Kolmogorov-Smirnov test. The time-bin dependence in the fluorescence trajectory on the probability density is carefully analyzed. We also discuss the ON and OFF processes from failure-rate analysis used in life testing of semiconductor devices.
\end{abstract} 

\maketitle

\section{Introduction}\label{sec_Introduction}
Materials science relating to fluorescence phenomena has long been one of the most active research fields. The luminescent properties of quantum dots are expected to have many applications~\cite{Zhuang_1998,Pirandola_2015,Huffaker_1998,Nozik_2002,Imamog_1999} due to the variable fluorescence as a function of the system size and a high fluorescence quantum yield. In addition, fluorescently labeled DNA, in which a fluorescent dye is embedded to a specific DNA sequence, has become an essential tool in molecular biochemistry~\cite{Kapuscinski_1995}, biomedical research~\cite{Xiaojun_2003}, and biophysics~\cite{Ha_2012} to understand DNA dynamics and interactions. 

\begin{figure}[b]
\centering
\includegraphics[width=\linewidth]{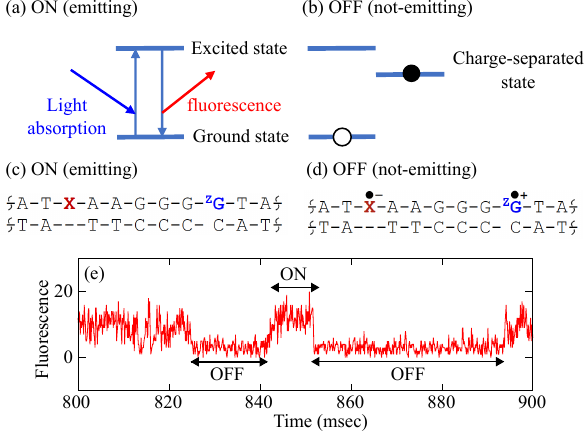}
\caption{A schematic figure of fluorescence blinking. When illuminated with light, a fluorophore takes an ON state [panel (a)], where it repeats light-absorbing and light-emitting. The system also takes an OFF state [panel (b)], where the light-emitting is interrupted because of the transition to not-emitting state. Panels (c) and (d) are schematic chemical structures of the fluorescently labeled DNA (double helix structure) to be analyzed~\cite{Fan_2022_2, Fan_2022_1}. The symbols A, G, T, and C are the bases that make up DNA, and X is the fluorescent molecule ATTO655, and $^{{\rm Z}}$G is a hole-trapping molecule. The panel (c) represents the ON state, and (d) represents the OFF state in which the molecule is trapped in a charge-separated state and does not emit light. See Sec.~\ref{DNA} for more  details. The panel (e) is a part of the observed fluorescence trajectory in Fig.~\ref{time_series_data} (a) of the present fluorecently labeled DNA, in which we see noticeable noises.}
\label{schmmatic}
\end{figure}

One of characteristic and important aspects in the florescent property is fluorescence blinking~\cite{Galland_2011,Sabanayagam_2005}. The state in which a fluorescent material repeatedly absorbs light and emits fluorescence under irradiation is called the ON state displayed in Fig.~\ref{schmmatic}(a), and the state in which it does not emit light is called the OFF state in Fig.~\ref{schmmatic}(b). When the fluorescence intensity is monitored as a function of time, the ON and OFF states are repeated, called fluorescence blinking, and the time series data shown in Fig.~\ref{schmmatic}(e) are generated~\cite{Fan_2022_2, Fan_2022_1}. This time series record the duration of the ON and OFF states. 
The blinking plot is obtained by calculating a probability density function of the duration of each states, providing a basis to understand the luminescent properties of materials. It should be noted here that the fluorescence trajectory contains a large amount of noise such as light emission from the substrate. When the noise becomes larger than the fluorescent signal, it becomes difficult to distinguish between the ON and OFF states, and the quantitative nature of the evaluated duration is ambiguous. As a result, the reliability of the blinking plot is also lost. A data analysis method that can remove noise and stably determine the ON/OFF states is desired.

In the present study, we report a hidden Markov model (HMM) analysis for the fluorescence trajectory of a fluorescently labeled DNA. The HMM is a machine learning method to handle time series data, and has been applied to various dynamic data such as stock price prediction~\cite{HASSAN_2009} and anomaly detection~\cite{KWON_1999}. In materials science, there are several applications to quantum dots~\cite{Furuta_2022}, random telegraph noise~\cite{Puglisi_2013,Puglisi_2014_1,Puglisi_2014_2,Stampfer_2018,Alok_2022}, nitrogen-vacancy centers in diamond~\cite{Murai_2018,Doi_2016}, electron holograms~\cite{Tamaoka_2021}, electron holography~\cite{Lee_2024}, atom quantum jump dynamics~\cite{Gammelmark_2014}, and crack propagation~\cite{Nguyen-Le_2020}. The fluorescent trajectory is a typical stochastic process including noise, and it is important to understand the physics behind this process. The HMM introduces hidden variables characterizing states behind the observed data. The ON/OFF states can be identified with a time series of the hidden variables instead of real fluorescence trajectory. An important point here is that the time series of the hidden variable are noise-suppressive, and as a result, we can stably estimate ON/OFF duration and the blinking plot.

On the study about a fluorescently labeled single molecule, fluctuation and correlation in the fluorescent intensity are important objects, because they include details of the composition, shape, diffusion, dynamics, intramolecular interaction, and environment effects~\cite{Orrit_2023}. Methods based on the auto correlation function~\cite{Lippitz_2005} and photon counting histogram~\cite{Hajdziona_2009} to the time series data play an important role. There are many reports with these techniques, including conformational dynamics of a single protein~\cite{Weixiang_2018}, enzymatic turnovers of single flavoenzyme molecules~\cite{Lu_1998}, electron transfer reaction of azurin molecules~\cite{Biswajit_2020}. HMM analyses have also been applied to the analyses of fluorescence resonance energy transfer~\cite{McKinney_2006,Pirchi_2016,Lee_2009}, diffusion of nanoparticles in the cytoplasm~\cite{Janczura_2021}, biomolecular motors~\cite{Okazaki_2020}, multichromophore photobleaching of dextran polymers~\cite{Messina_2006}. In the present study, we apply the HMM analysis to the understanding of fluorescence blinking of the fluorescently labeled DNA. 

This paper is organized as follows: In Sec.~\ref{sec_method}, we describe an experimental setup for single-molecule measurements of fluorescent signals of a fluorecently labeled DNA, and overviews of the HMM analysis and how to calculate the blinking plot. In Sec.~\ref{sec_results_and_discussions}, we show results of the HMM analysis for the experimental fluorescence trajectory and model fitting to the probability density for the duration of the ON and OFF states. The best fitting functions are pursued in term of the statistical hypothesis testing of goodness-of-fit. In addition, we perform a failure-rate analysis to understand the ON$\to$OFF and OFF$\to$ON transition processes. Section~\ref{sec_summary} summarizes conclusions.

\section{Method}\label{sec_method} 

\subsection{Single molecule measurement for fluorescence trajectory}~\label{DNA}
In the present study, we analyze the fluorescence fluctuation data previously measured for charge-separation and charge-recombination processes in DNA~\cite{Fan_2022_2, Fan_2022_1}. Figures~\ref{schmmatic} (c) and (d) are schematic chemical structures of the fluorescently labeled DNA (double helix structure) to be analyzed, where A, G, T, and C are the bases that make up DNA, X is the fluorescent molecule ATTO655 as a photosensitizer, and $^{{\rm Z}}$G is a deazaguanine as a hole trap to observe a charge-separation and charge recombination process in DNA at the single-molecule level. The panel (c) represents a schematic chemical structure for an emitting ON state, where ATTO655 repeats photon-absorption and emission. The panel (d) represents a schematic chemical structure of a not-emitting OFF state; during the charge-separated state, ATTO655 is in the radical anion form and cannot emit, therefore the duration of the OFF state corresponds to the lifetime of the charge-separated state. 

Briefly, for the experimental setup, biotin group was introduced at the end of the duplex and was anchored on a glass surface through well-established biotinylated-BSA, streptavidin, biotinylated-DNA chemistry. The single molecule measurement was performed on a custom-made confocal fluorescence microscope consisting of an inverted optical microscope (IX73, Olympus). The output of the laser (OBIS 637 LX, Coherent, 637 nm, 4.6 $\mu$W) was focused in the sample by an objective (UPlanXApo, x60, oil, NA 1.42, Olympus), and the detection volume (confocal volume) was regulated by a pinhole (diameter 25 $\mu$m, Thorlabs MPH16). The scattered light was blocked by a band-pass filter (FF01-697/58-25, Semrock), and the emitted photons from ATTO655 were detected by an avalanche photodiode (SPCM-AQRH-14, Perkin-Elmer). A counting board (SPC-130EMN, Becker \& Hickl GmbH) was used to count the output, and real-time monitoring of fluorescence intensity fluctuations has been achieved using the Spcm64 system. (Becker \& Hickl GmbH). The time resolution of this device on the photon counting is 80 nsec. 

\subsection{Hidden Markov model}\label{Hidden Markov model}
We next briefly describe our calculation method of HMM~\cite{Furuta_2022}. In the HMM, a probability density function combined with various stochastic variables is introduced. The joint distribution of the HMM employed in the present study is the following form as
\begin{eqnarray}
 p({\bm I}, \mathbf{S}, \mathbf{\Theta}, \bm{\pi}, \mathbf{A}) 
 =p({\bm I}|\mathbf{S}, \mathbf{\Theta}) 
 p(\mathbf{S}|\bm{\pi},\mathbf{A}) 
 p(\mathbf{\Theta}) 
 p(\bm{\pi}) 
 p(\mathbf{A}). \nonumber \\
 \label{douji-2}
\end{eqnarray}
Here, 
\begin{eqnarray}
 {\bm I}=(I_1, I_2, \cdots, I_N) \label{I_trajectory}
\end{eqnarray} 
is observed data corresponding to a fluorescence trajectory to be analyzed, and $N$ is the total numbers of the time grids. $\mathbf{S}$ is a time series of hidden variables as 
\begin{eqnarray}
  \mathbf{S} = ({\bf s}_1, {\bf s}_2, \cdots, {\bf s}_N),  \label{S_trajectory}
\end{eqnarray} 
which describes the internal state of each time in the fluorescence trajectory, and ${\bf s}_i$ is a $K$-dimensional vector in the time step $i$. To identify an ON or OFF state, we set $K$ = 2~\cite{Furuta_2022}. $\mathbf{\Theta}$ describes the parameter characterizing a distribution function of observed data ${\bm I}$. In the present study, we assume that the distribution of each state follows the Gaussian distribution with a mean $\mu$ and a precision $\lambda$, where $\lambda ^{-1}$ represents a variance. Then, we write $\mathbf{\Theta}$ as $\{{\bm \mu}, {\bm \lambda}\}$ with ${\bf \mu}=(\mu_{{\rm ON}}, \mu_{{\rm OFF}})$ and ${\bf \lambda}=(\lambda_{{\rm ON}}, \lambda_{{\rm OFF}})$. $\bm{\pi}$ represents the probability of the initial-step hidden variable, and $\bm{\pi} = (\pi_{{\rm ON}}, \pi_{{\rm OFF}})$. $\mathbf{A}$ describes a 2$\times$2 transition matrix for the time evolution of the hidden variables. $p(\mathbf{S}|\bm{\pi}, \mathbf{A})$ in Eq.~(\ref{douji-2}) describes the conditional probability distribution of $\mathbf{S}$ with $\mathbf{A}$ and $\bm{\pi}$ fixed, and similarly, $p(\bm{I}|\mathbf{S},\mathbf{\Theta})$ is the conditional probability distribution of $\bm{I}$ after $\mathbf{S}$ and $\mathbf{\Theta}$ were determined.

The purpose of the HMM simulation is to infer the hidden-variable time series $\mathbf{S}$. For this purpose, we need to calculate a posterior distribution which is a conditional distribution function under the observed trajectory 
$\bm{I}$ as
\begin{eqnarray}
  p(\mathbf{S},{\bm \mu}, {\bm \lambda}, \bm{\pi},  \mathbf{A}|\bm{I})
  =\frac{p(\bm{I}, \mathbf{S}, \bm{\mu}, \bm{\lambda}, \bm{\pi}, \mathbf{A})}{p(\bm{I})}, 
  \label{conditional-p}
\end{eqnarray}
where $p(\bm{I})$ is the marginal distribution written as
\begin{eqnarray}
  p(\bm{I})
  =\sum_{\mathbf{S}} \int p(\bm{I}, \mathbf{S}, \bm{\mu}, \bm{\lambda}, \bm{\pi}, \mathbf{A}) d\bm{\mu} d\bm{\lambda} d\bm{\pi} d\mathbf{A}. 
  \label{pI}
 \end{eqnarray}
To calculate the posterior distribution in Eq.~(\ref{conditional-p}) approximately, we use a blocking Gibbs sampling based on Bayesian inference. Optimization details can be found in Ref.~\onlinecite{Furuta_2022}. By evaluation of the posterior distribution, the hidden-variable time series $\mathbf{S}$ in Eq.~(\ref{S_trajectory}) can be obtained as a simulation result. On the calculation condition, the total number of the iteration steps of the Gibbs sampling $N_{itr}=1000$, and the hyperparameters of each distribution are set to the same as Ref.~\onlinecite{Furuta_2022}.

\subsection{Probability density of duration} \label{sec_experimental_data_histogram}
Next, we describe how to analyze the hidden-variable time series $\mathbf{S}$ in Eq.~(\ref{S_trajectory}) obtained from the HMM simulation. For this purpose, we consider a probability density of duration $\tau$ as 
\begin{align}
    p(\tau)=\frac{1}{N_e} 
    \sum_{\alpha =1}^{N_{e}}
    \delta(\tau-\tau_{\alpha}), 
    \label{p_conti}
\end{align}
where $p(\tau)$ is normalized as $\int p(\tau) d\tau=1$. We consider $p(\tau_{{\rm ON}})$ and $p(\tau_{{\rm OFF}})$, and $\tau_{\alpha}$ in Eq.~(\ref{p_conti}) is the $\alpha$-th ON or OFF event duration. The \{$\tau_{\alpha}$\} data are collected from the hidden-variable time series $\mathbf{S}$~\cite{Furuta_2022}, and $N_{e}$ is the total number of the \{$\tau_{\alpha}$\} data. In the practical calculation, $p(\tau)$ is evaluated discretely as
\begin{align}
    p(\tau_{k})=\frac{1}{C}
    \sum_{\alpha =1}^{N_{e}}
    \delta(\tau_{k}-\tau_{\alpha}),  
    \label{p_hist}
\end{align}
where $p(\tau_{k})$ is proportional to the total number of the events with the duration from $\tau_{k-1}$ to $\tau_k$, and $\tau_k$ is given as $k\Delta_{\tau}$ with $\Delta_{\tau}$ being a grid spacing of duration. $C$ is a normalization constant given as 
\begin{align}
    C=
    \sum_{k =1}^{N_{\tau}} \sum_{\alpha =1}^{N_{e}}
    \delta(\tau_{k}-\tau_{\alpha})
     \Delta\tau, 
    \label{p_sum}
\end{align}
where $N_{\tau}$ is the total number of the duration grids.  

The calculation condition is as follows: We analyze the fluorescence trajectories of 40 molecules. The length of the fluorescence trajectories is ranging from 0.46 sec to 9.27 sec. For the time bin $\Delta$ of the trajectory, we consider the three size; 62.5 $\mu$sec, 125 $\mu$sec, and 250 $\mu$sec. The total number of the time grids of each trajectory, $N$ in Eqs.~(\ref{I_trajectory}) and (\ref{S_trajectory}), depends on the trajectory length and $\Delta$. The total number of the \{$\tau_{\alpha}$\} data, $N_e$ in Eq.~(\ref{p_conti}) or (\ref{p_hist}), is 2,078 for the ON case and 2,067 for the OFF case with the $\Delta$=250-$\mu$sec case. Similarly, for the $\Delta$=125-$\mu$sec case, $N_e$ is 2,341 (ON) and 2,334 (OFF), and for the $\Delta$=62.5-$\mu$sec case,  $N_e$ is 2,697 (ON) and 2,678 (OFF). The grid spacing of duration $\Delta\tau$ is 500 $\mu$sec, and the total number of the duration grids $N_{\tau}$ is 462 for the ON case and 122 for the OFF case with the $\Delta$=250-$\mu$sec case. Similarly, for the $\Delta$=125-$\mu$sec case, $N_{\tau}$ is 383 (ON) and 122 (OFF), and for the $\Delta$=62.5-$\mu$sec case, $N_{\tau}$ is  322 (ON) and 122 (OFF).

\section{Results and Discussions}\label{sec_results_and_discussions} 

\subsection{Analysis for fluorescence trajectory}  
\begin{figure*}[!]
\begin{center}
\includegraphics[width=0.9\linewidth]{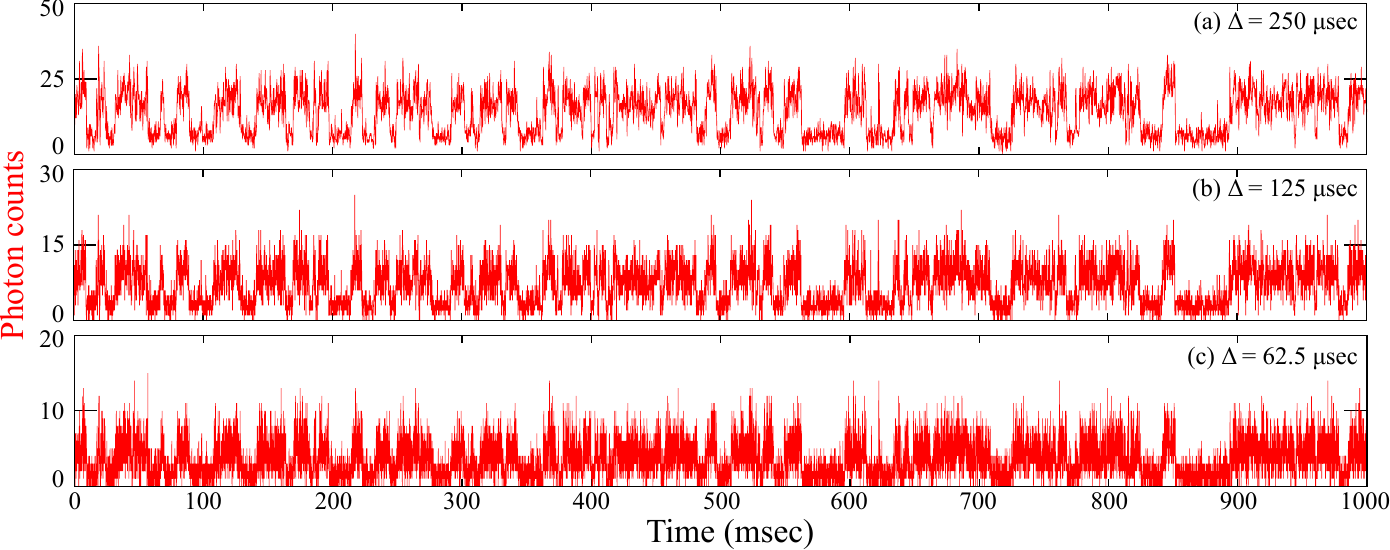}
\end{center}
\caption{Three experimental photon-count trajectories corresponding to Eq.~(\ref{I_trajectory}), where the photon counts are evaluated in the time-bin $\Delta$: (a) $\Delta$ = 250 $\mu$sec, (b) $\Delta$ = 125 $\mu$sec, and  (c) $\Delta$ = 62.5 $\mu$sec.}
\label{time_series_data}
\end{figure*}

Figure~\ref{time_series_data} shows three experimental fluorescence trajectories corresponding to Eq.~(\ref{I_trajectory}), where photon counts are evaluated in the time bin $\Delta$ = (a) 250 $\mu$sec, (b) 125 $\mu$sec, and (c) 62.5 $\mu$sec.  These data are all the same trajectory, but $\Delta$ is different, so the integrated values for each bin are different. It can clearly be seen that the noise becomes more appreciable as the $\Delta$ becomes smaller. 

\begin{figure*}[!]
\centering
\includegraphics[width=0.94\linewidth]{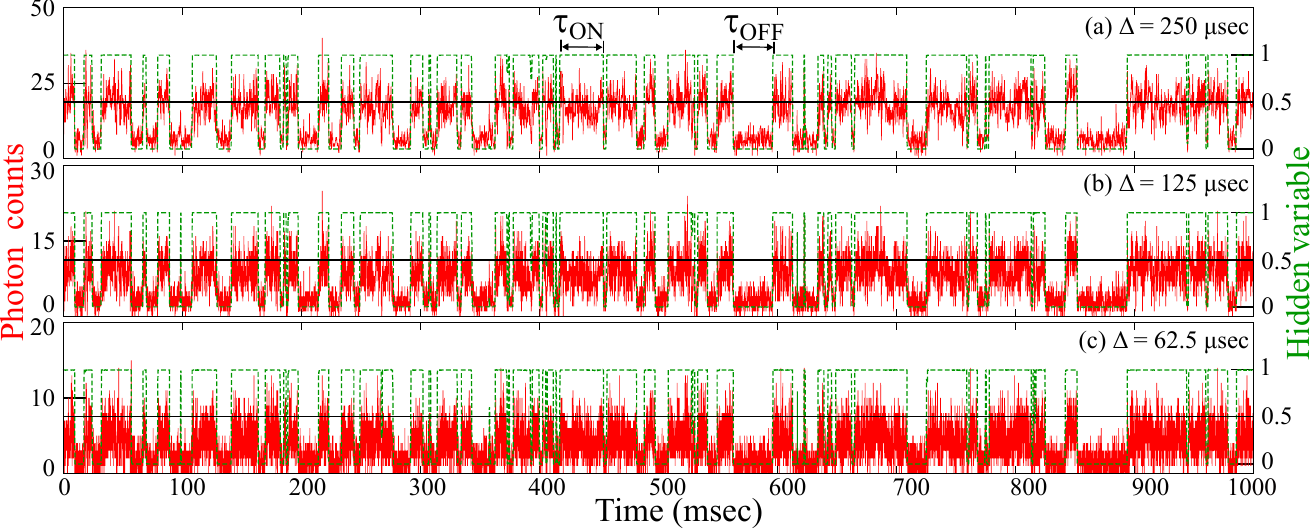}
\caption{Three trajectories of hidden variables $\mathbf{S}$ in Eq.~(\ref{S_trajectory}) obtained from HMM simulations (green-dashed lines): (a) $\Delta$ = 250 $\mu$sec, (b) $\Delta$ = 125 $\mu$sec, and  (c) $\Delta$ = 62.5 $\mu$sec. Black-solid lines of 0.5 are drawn to distinguish an ON (above 0.5) state and an OFF (below 0.5) state for  $\mathbf{S}$. Examples of the ON and OFF duration, $\tau_{{\rm ON}}$ and $\tau_{{\rm OFF}}$, are illustrated in the panel (a). Red-solid lines are the photon-count trajectories and how to see is the same as Fig.~\ref{time_series_data}.}
\label{HMM}
\end{figure*}

Figure~\ref{HMM} superimposes the hidden-variable time series $\mathbf{S}$ in Eq.~(\ref{S_trajectory}) obtained from the HMM simulations, denoted by the green dashed lines, onto the original fluorescent trajectories (red solid lines). A black-solid line of 0.5 is the line to distinguish an ON state (above 0.5) and an OFF state (below 0.5) for the hidden variable time series. In the practical calculation, this threshold is slightly adjusted depending on the data within 0.4-0.6. As can be seen from the figure, the hidden-variable time series well capture the switching of the ON$\to$OFF and OFF$\to$ON process, from which we properly evaluate the ON and OFF duration, $\tau_{{\rm ON}}$ and $\tau_{{\rm OFF}}$. Examples of $\tau_{{\rm ON}}$ and $\tau_{{\rm OFF}}$ are illustrated in the Fig.~\ref{HMM} (a). Since the extremely noisy data affects the state identification, short-duration events might slightly be overestimated in the analysis of time series with smaller $\Delta$. We note that the moving-average method~\cite{Ellis_2005} does not work for the ON/OFF identification from the present fluorescence trajectories. This is because the data is noisy, so there is a limitation for data smoothing. As a result, a large number of artificial short-duration events are generated, compared to the HMM. In this sense, the HMM is highly effective for the noise removing.

\subsection{Probability density of duration}~\label{PON_and_POFF} 
\begin{figure}[!]
\centering
\includegraphics[width=\linewidth]{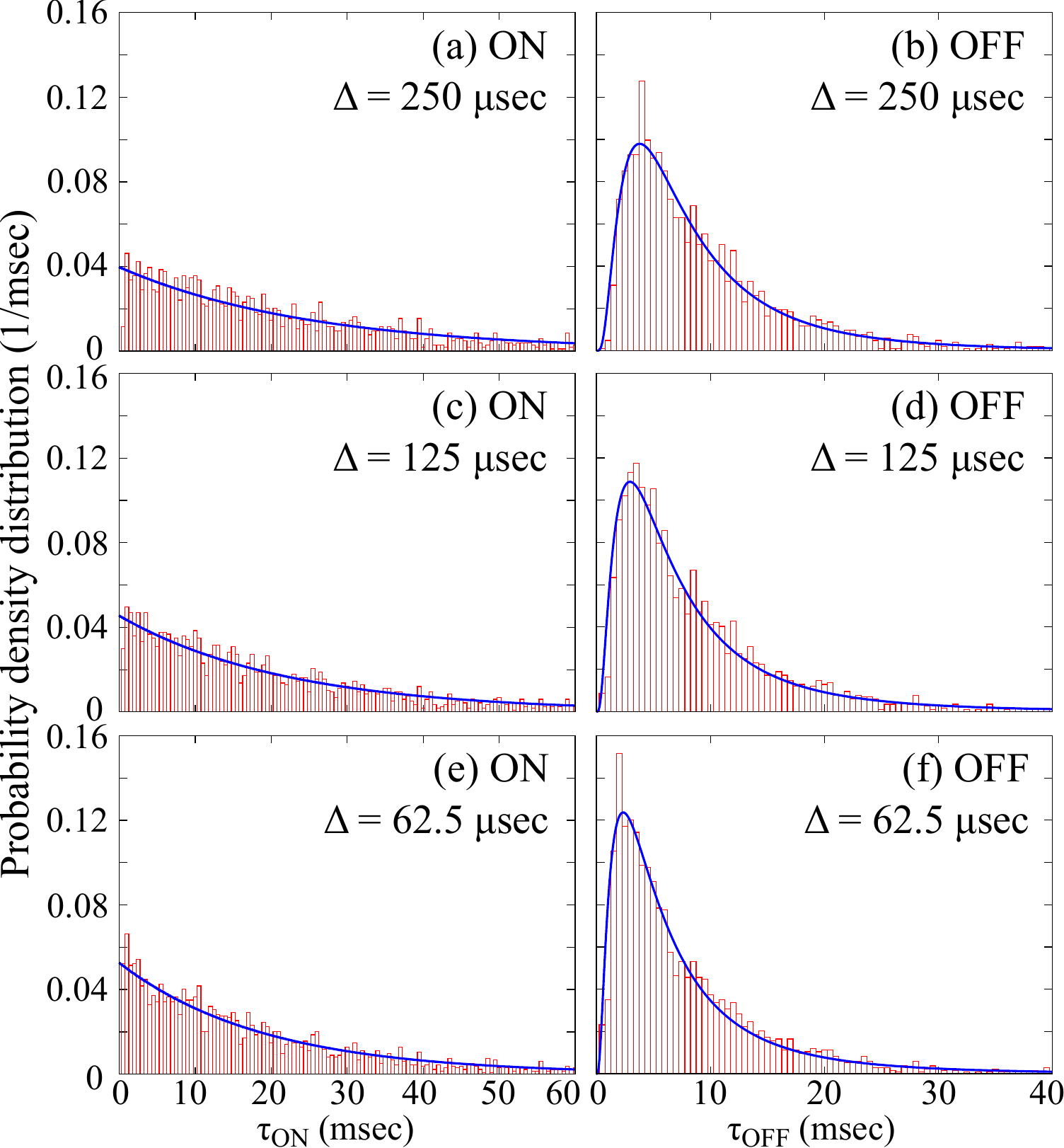}
\caption{Our calculated probability density for the ON/OFF duration in fluorescence blinking, $p(\tau_{{\rm ON}})$ and $p(\tau_{{\rm OFF}})$ in Eq.~(\ref{p_hist}), denoted by red-bars. Blue-solid curves represent exponential distributions $P(\tau_{{\rm ON}})$ in Eq.~(\ref{expon}) and log-normal distributions $P(\tau_{{\rm OFF}})$ in Eq.~(\ref{lognorm}), whose optimized parameters are summarized in Table~\ref{table_ON_OFF}. The upper (a) and (b), middle (c) and (d), and lower (e) and (f) panels are the results for the time series with $\Delta$ = 250 $\mu$sec, 125 $\mu$sec, and 62.5 $\mu$sec, respectively. Also, the left (a), (c), and (e) panels are results for the ON duration, while the right (b), (d), and (f) panels correspond to the results for the OFF duration.}
\label{experimental_data_histogram}
\end{figure}

Figure~\ref{experimental_data_histogram} displays our calculated probability density for the ON [(a), (c), and (e)] and OFF [(b), (d), and (f)] duration, $p(\tau_{k})$ in Eq.~(\ref{p_hist}). The upper (a) and (b) describe the results for fluorescence trajectories with $\Delta$ = 250 $\mu$sec. The middle (c) and (d) and the lower (e) and (f) panels describe the results for $\Delta$ = 125 $\mu$sec and $\Delta$ = 62.5 $\mu$sec, respectively. Note that these data are collected from the fluorescence trajectories of the 40 molecules. 

To analyze the trends of these data, we performed fittings of model probability density functions. For the ON distribution, we use an exponential distribution as
\begin{align}
    P(\tau_{{\rm ON}})=\frac{1}{s}\exp \biggl( - \frac{\tau_{{\rm ON}} }{s} \biggr), 
    \label{expon}
\end{align}
where $s$ is a parameter for the exponential function. For the OFF distribution, we consider a log-normal distribution as
\begin{align}
    P(\tau_{{\rm OFF}})=\frac {1}{\sqrt{2\pi \sigma^2} \tau_{{\rm OFF}}} \exp \biggl( -\frac {(\ln \tau_{{\rm OFF}} -\mu)^2}{2 \sigma^2 } \biggr),
\label{lognorm}
\end{align}
where $\mu$ and $\sigma$ represent mean and standard deviation of the distribution, respectively. To evaluate the fitting accuracy, we calculate the following measures; root mean squared error (RMSE) as 
\begin{align}
    {\rm RMSE}=\sqrt{\frac{1}{N_{D}}
    \sum_{k=1}^{N_{D}}
    \bigl(p(\tau_{k})-P(\tau_{k})\bigr)^2},  
    \label{RMSE}
\end{align}
and coefficient of determination ${\rm R}^2$ as
\begin{align}
    {\rm R}^2= 1 - \frac{\sum_{k=1}^{N_{D}}
    \bigl( p(\tau_{k})-P(\tau_{k}) \bigr)^2} 
    {\sum_{k=1}^{N_{D}} \bigl( p(\tau_{k})- \bar{p} \bigr)^2} 
    \label{R2}
\end{align}
with the $\bar{p}$ in Eq.~(\ref{R2}) being the mean of probability density as
\begin{align}
    \bar{p}=\frac{1}{N_{D}} \sum_{k=1}^{N_{D}} p(\tau_{k}).
    \label{pbar}
\end{align}
Here, $\tau_k$ represents the $k$-th duration grid, introduced in Eq.~(\ref{p_hist}), but the calculations of Eqs.(\ref{RMSE}), (\ref{R2}), and (\ref{pbar}) exclude the data of $p(\tau_k)=0$. Thus, the total number of the data points $N_{D}$ is 231 for the ON case and 88 for the OFF case with the $\Delta$=250-$\mu$sec case. Similarly, for the $\Delta$=125-$\mu$sec case, $N_{D}$ is 218 (ON) and 85 (OFF), and for the $\Delta$=62.5-$\mu$sec case, $N_{D}$ is  209 (ON) and 87 (OFF). The $P(\tau_{k})$ is the value of the fitting function of the $k$-th duration grid. 

The model parameters in Eqs.~(\ref{expon}) and (\ref{lognorm}) are determined as follows: We first perform least-squared fitting. Then, we set the resulting parameters to the initial guess, and perform the Kolmogorov-Smirnov (KS) test~\cite{Massey_1951}. In this test, we tune the model parameters to maximize $p$-value for measuring goodness-of-fit, where the $p$-value is defined as 
\begin{eqnarray}
 p\mathchar`-{\rm value}=p(z)=2\sum_{j=1}^{\infty}(-1)^{j-1}\exp(-2j^2z^2) 
 \label{pval}
\end{eqnarray}
with
\begin{eqnarray}
z=\sqrt{N_e} \max_{\tau} \bigl| S_{N_e}(\tau)-F(\tau) \bigr|
\end{eqnarray}
and $N_e$ being the total number of the \{$\tau_{\alpha}$\} data [Eq.~(\ref{p_conti})]. $S_{N_e}(\tau)$ is an empirical distribution function as 
\begin{eqnarray}
   S_{N_e}(\tau) = \frac{1}{N_e}\sum_{\alpha=1}^{N_e} {\bf 1}_{\tau_{\alpha}<\tau}, 
\end{eqnarray}
where ${\bf 1}_{A}$ is an indicator function for the condition $A$, and $F(\tau)$ is a cumulative distribution function as 
\begin{eqnarray}
    F(\tau)=\int_{0}^{\tau} P(\tau') d\tau', 
\end{eqnarray}
where $P(\tau)$ is a probability density function to be tested. The conventional significance level for the $p$-value is 0.05, and if the obtained $p$-value is above 0.05, the test model function is considered to have validity. 

Figure~\ref{experimental_data_histogram} shows the resulting fitted curves with blue-solid lines. We see from the figure that the fittings are satisfactorily. Optimized parameters and fitting accuracy are given in Table~\ref{table_ON_OFF}, where the first and second columns contain the results for $P(\tau_{{\rm ON}})$ and $P(\tau_{{\rm OFF}})$, respectively. From the table, the RMSE and ${\rm R}^2$ support good accuracy; very low RMSE and ${\rm R}^2$ sufficiently close to 1. The resulting $p$-values are also listed, where they are 0.2-0.6 for $P(\tau_{{\rm ON}})$ and nearly 0.3 for $P(\tau_{{\rm OFF}})$. These $p$-values are larger than the conventional significance level of 0.05. Thus, the null hypothesis (the data distribution differs from the assumed distribution function) is rejected.
\begin{table*}[!]
\caption{Summary of fittings: We employ an exponential function $P(\tau_{{\rm ON}})$ in Eq.~(\ref{expon}) for the fitting of the ON-duration probability densities in Figs.~\ref{experimental_data_histogram} (a), (c), and (e), and a log-normal function $P(\tau_{{\rm OFF}})$ in Eq.~(\ref{lognorm}) for the OFF-duration probability density in Figs.~\ref{experimental_data_histogram} (b), (d), and (f).  The $s$ is a parameter of $P(\tau_{{\rm ON}})$ in Eq.~(\ref{expon}), and the $\mu$ and $\sigma$ are parameters of $P(\tau_{{\rm OFF}})$ in Eq.~(\ref{lognorm}). RMSE and R$^2$ are calculated from Eq.~(\ref{RMSE}) and Eq.~(\ref{R2}), respectively. The $p$-value with the KS test in Eq.~(\ref{pval}) are listed. The fitting accuracy of the normal distribution $P(\alpha_{{\rm OFF}})$ in Eq.~(\ref{normal}) of Appendix~\ref{app:normal} to the data of Fig.~\ref{Fig_normal} is also shown, with $\alpha_{{\rm OFF}}=\ln\tau_{{\rm OFF}}$. It is noted that the parameters $\mu$ and $\sigma$ of the normal distribution are the same as those of the log-normal distribution, and the $p$-value of $P(\alpha_{{\rm OFF}})$ is the same as that of $P(\tau_{{\rm OFF}})$.} 
\begin{center}
\begin{tabular}{r@{\ \ \ }c@{\ \ \ }c@{\ \ \ }c@{\ \ \ }c@{\ \ \ }c@{\ \ \ }c@{\ \ \ }c@{\ \ \ }c@{\ \ \ }c@{\ \ \ }c@{\ \ \ }c@{\ \ \ }c@{\ \ \ }c} \hline \hline \\ [-6pt] 
         & \multicolumn{4}{c}{$P(\tau_{{\rm ON}})$}  & & \multicolumn{5}{c}{$P(\tau_{{\rm OFF}})$}  & & \multicolumn{2}{c}{$P(\alpha_{{\rm OFF}})$} \\ [2pt] 
         \cmidrule{2-5}                                      \cmidrule{7-11}                                      \cmidrule{13-14}   
$\Delta$ & $s$ (ms) & RMSE    & R$^2$ & $p$-value & & $\mu$ & $\sigma$ & RMSE    &R$^2$  & $p$-value& & RMSE   & R$^2$ \\ [2pt] \hline \\ [-8pt]
250      & 25.2 & 0.00339 & 0.896 & 0.196       & & 1.95  & 0.787    & 0.00557 & 0.966 & 
0.306& & 0.0399 & 0.955 \\ [2pt] 
125      & 22.0 & 0.00291 & 0.937 & 0.618      & & 1.82  & 0.860   & 0.00601 & 0.964 & 0.275        & & 0.0381 & 0.950 \\ [2pt] 
62.5     & 19.0 & 0.00294 & 0.950 & 0.460       & & 1.68  & 0.908    & 0.00659 & 0.962 & 0.275        & & 0.0341 & 0.953 \\ [2pt] 
\hline \hline
\end{tabular} 
\end{center}
\label{table_ON_OFF}
\end{table*}

Next, we discuss the characteristic relaxation time in terms of the resulting model parameters. For the relaxation time of the ON$\to$OFF process, it is estimated as 20-25 msec (see $s$ in Table~\ref{table_ON_OFF}). In addition, on the characteristic relaxation time $e^{\mu}$ of the OFF$\to$ON process, we obtain $e^{1.95}=7.03$ msec for the $\Delta=250$-$\mu$sec case, $e^{1.82}=6.17$ msec for the $\Delta=125$-$\mu$sec case, and $e^{1.68}=5.37$ msec for the $\Delta=62.5$-$\mu$sec case. The estimate based on the auto-correlation function analysis~\cite{Fan_2022_1,Fan_2022_2} is nearly 5 msec, and thus, the present HMM estimates are consistent with the estimates of the correlation analyses. We note that the both relaxation times of the ON and OFF states become short as the time-bin size $\Delta$ of the fluorescence trajectory becomes small. This may reflect that the longer duration event is split into the shorter duration events with considering the smaller $\Delta$.

\subsection{Best fit for probability density of the OFF duration}  
\label{sec_weibull_gamma_lognorm}
The ON distribution is basically a monotonically decreasing, so the validity of the exponential function is identified visibly. On the other hand, the OFF distribution has an asymmetric structure, so it is necessary to carefully verify the validity of the log-normal function. To find the best function for probability density of the OFF duration, we investigate two other representative asymmetric distributions, the Weibull and Gamma distributions. The Weibull distribution $P_{W}(\tau)$ is written as 
\begin{eqnarray}
 P_{W}(\tau)=\frac{m}{\eta} \biggl( \frac{\tau}{\eta} \biggr)^{m-1} \exp \biggl \{ - \biggl (\frac{\tau}{\eta} \biggr )^{m} \biggr \},
 \label{Weibull}
\end{eqnarray}
where $m$ and $\eta$ are the shape and scale parameters of the distribution, respectively. Also, the Gamma distribution $P_{G}(\tau)$ is written as
\begin{eqnarray}
 P_{G}(\tau)=\frac{b^{-a}}{\Gamma (a)} \tau^{a-1} \exp \biggl( -\frac{\tau}{b} \biggr ), 
 \label{Gamma}
\end{eqnarray}
where $a$ and $b$ are the shape and scale parameters of the distribution, respectively. $\Gamma (a)$ is the Gamma function.

\begin{figure*}[!]
\centering
\includegraphics[width=0.65\linewidth]{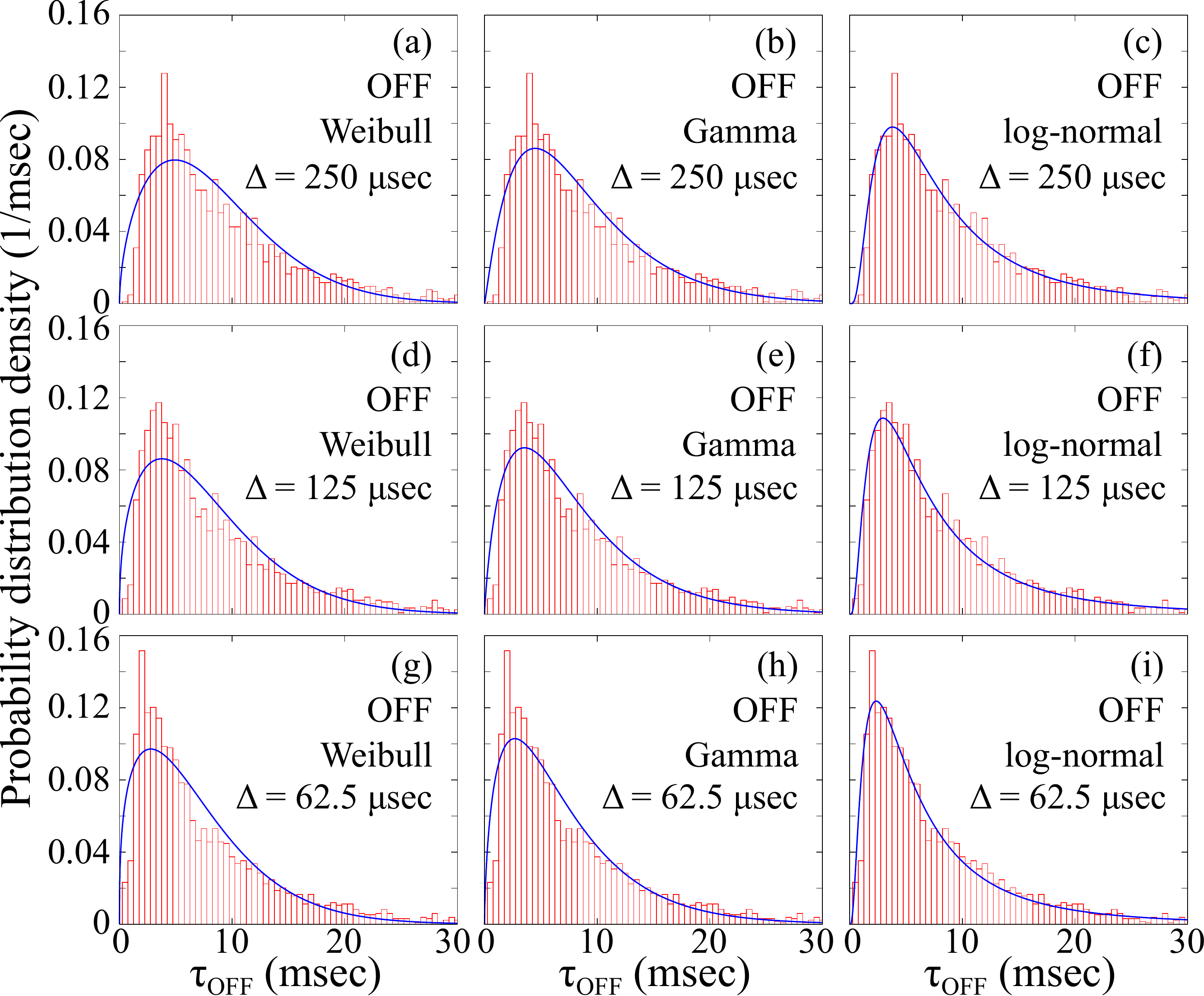}
\caption{Comparison of fittings to the OFF-duration probability density in Fig.~\ref{experimental_data_histogram}: Panels (a), (d), and (g) are the fitting results with the Weibull distribution $P_{W}(\tau_{{\rm OFF}})$ in Eq.~(\ref{Weibull}), and (b), (e), and (h) are the Gamma distribution $P_{G}(\tau_{{\rm OFF}})$ in Eq.~(\ref{Gamma}). The panels (c), (f), and (i) are the log-normal distribution $P(\tau_{{\rm OFF}})$ in Eq.~(\ref{lognorm}). Also, (a), (b), and (c) describe the results for the time bin $\Delta=250$ $\mu$sec, and (d), (e), and (f) describe the results for $\Delta=125$ $\mu$sec. The (g), (h), and (i) are the results for $\Delta=62.5$ $\mu$sec. The fits are given with blue-solid curves, and fitted parameters are given in Table~\ref{table_Weibull_Gamma_Log-norm}. } 
\label{Weibull_Gamma_Log-norm}
\end{figure*}

Figure~\ref{Weibull_Gamma_Log-norm} compares the fittings to the OFF probability density: The panels (a), (d), and (g) are the fitting results of the Weibull distribution $P_{W}(\tau_{{\rm OFF}})$ in Eq.~(\ref{Weibull}), based on the least-squared fitting plus the KS-$p$ value maximization, and (b), (e), and (h) are the fitting results of the Gamma distribution $P_{G}(\tau_{{\rm OFF}})$ in Eq.~(\ref{Gamma}). The panels (c), (f), and (i) are the log-normal distribution $P(\tau_{{\rm OFF}})$ in Eq.~(\ref{lognorm}). Also, (a), (b), and (c) describe the results for the time bin $\Delta=250$ $\mu$sec, (d), (e), and (f) describe the results for $\Delta=125$ $\mu$sec, and (g), (h), and (i) are the results for $\Delta=62.5$ $\mu$sec. The fitted parameters are summarized in Table~\ref{table_Weibull_Gamma_Log-norm}. From the figure, we see that the log-normal function gives a good fit, while the Weibull and Gamma functions seem to underestimate the peak height around $\tau_{{\rm OFF}}\sim$ 5 msec. In the Weibull and Gamma functions, the tail behavior in the low duration side is not described properly. The resulting $p$-values for the Weibull and Gamma distributions are appreciably small compared to those of the log-normal distribution, and well below the significant level of 0.05. 
\begin{table*}[!]
\caption{Summary of fitting functions to the OFF-duration probability density: The fitting functions are the Weibull $P_{W}(\tau_{{\rm OFF}})$ in Eq.~(\ref{Weibull}), Gamma $P_{G}(\tau_{{\rm OFF}})$ in Eq.~(\ref{Gamma}), and log-normal $P(\tau_{{\rm OFF}})$ in Eq.~(\ref{lognorm}). The fitted curves are plotted in Fig.~\ref{Weibull_Gamma_Log-norm}. The $m$ and $\eta$ are the fitted parameters of $P_{W}(\tau_{{\rm OFF}})$, and the $a$ and $b$ are the fitted parameters of $P_{G}(\tau_{{\rm OFF}})$. Also, the $\mu$ and $\sigma$ are fitted parameters of $P(\tau_{{\rm OFF}})$. Evaluated $p$-values with the KS test in Eq.~(\ref{pval}) are listed.} 
\begin{center}
\begin{tabular}{c@{\ \ \ }
c@{\ \ \ }c@{\ \ \ }c@{\ \ \ }c@{\ \ }
c@{\ \ \ }c@{\ \ \ }c@{\ \ \ }c@{\ \ }
c@{\ \ \ }c@{\ \ \ }c} \hline \hline \\ [-6pt] 
 & \multicolumn{3}{c}{Weibull} & & \multicolumn{3}{c}{Gamma} & & \multicolumn{3}{c}{log-normal} \\ [2pt] 
   \cmidrule{2-4} \cmidrule{6-8} \cmidrule{10-12}   
$\Delta$ & $m$   &$\eta$ (ms)  & $p$-value & & 
           $a$   &$b$ (ms)     & $p$-value & & 
           $\mu$ & $\sigma$         & $p$-value \\ [2pt] \hline \\ [-8pt]
250  & 1.56 & 9.47 & 4.32$\times 10^{-5}$ & & 
       2.09 & 4.12  & 3.10$\times 10^{-3}$ & & 
           1.95  & 0.787      & 0.306 \\ [2pt] 
125  & 1.44 & 8.57  & 1.82$\times 10^{-4}$ & & 
       1.82 & 4.33  & 6.81$\times 10^{-3}$ & & 
           1.82  & 0.860      & 0.275 \\ [2pt] 
62.5 & 1.35 & 7.56  & 7.59$\times 10^{-5}$ & & 
       1.62 & 4.34  & 1.95$\times 10^{-3}$ & & 
           1.68  & 0.908      & 0.275 \\ [2pt] 
\hline \hline
\end{tabular} 
\end{center}
\label{table_Weibull_Gamma_Log-norm}
\end{table*}

The fact that $\tau_{{\rm OFF}}$ follows the log-normal distribution means that $\tau_{{\rm OFF}}$ is based on the multiplicative stochastic process as~\cite{Kolmogorov_1941,Ishii_1992,Antoniou_2002}  
\begin{eqnarray}
    \tau_{{\rm OFF}} = \prod_{i=1}^{M} n_i, 
    \label{tauoff}
\end{eqnarray}
where $n_i$ is the $i$-th factor affecting $\tau_{{\rm OFF}}$ and $M$ is the total numbers of the factors. Now, taking the logarithm of Eq.~(\ref{tauoff}), we obtain  
\begin{eqnarray}
    \ln \tau_{{\rm OFF}} = \sum_{i=1}^{M} \ln n_i 
    \label{lntauoff} 
\end{eqnarray}
If $n_i$ is independent with each other and $M$ is sufficiently large, $\ln \tau_{{\rm OFF}}$ follows the normal distribution based on the central limit theorem, or equivalently, $\tau_{{\rm OFF}}$ follows the log-normal distribution. Now, an important question is what is the factor $n_i$ affecting the stability of the OFF state, leaving to be explored.

\subsection{Failure rate analysis}\label{sec_Failure_rate}
Using the above determined model distribution functions, we discuss aspects of the ON$\to$OFF and OFF$\to$ON transition processes. The ON$\to$OFF transition corresponds to a process in which the photon-absorption and emission cycle changes to a charge-separated state. On the other hand, the OFF$\to$ON transition corresponds to a process in which the charge-separated state returns to the ground state. In addition, it was confirmed that the lifetime distribution of the ON state decays exponentially, and that of the OFF state exhibits the log-normal distribution.

Based on these results, we will discuss aspects of the transition processes in terms of the failure analysis in the semiconductor device~\cite{Klutke_2003}. In this analysis, for example, the ON$\to$OFF process is discussed as follows: First, an ON state is defined as a non-faulty state, and an OFF state is defined as a faulty state. The failure analysis tells us the nature of the failures; we can classify whether the ON$\to$OFF process is based on an accidental event or a wear-out event, as well as the failure type of the semiconductor product. 

\begin{figure}[!]
\centering
\includegraphics[width=0.9\linewidth]{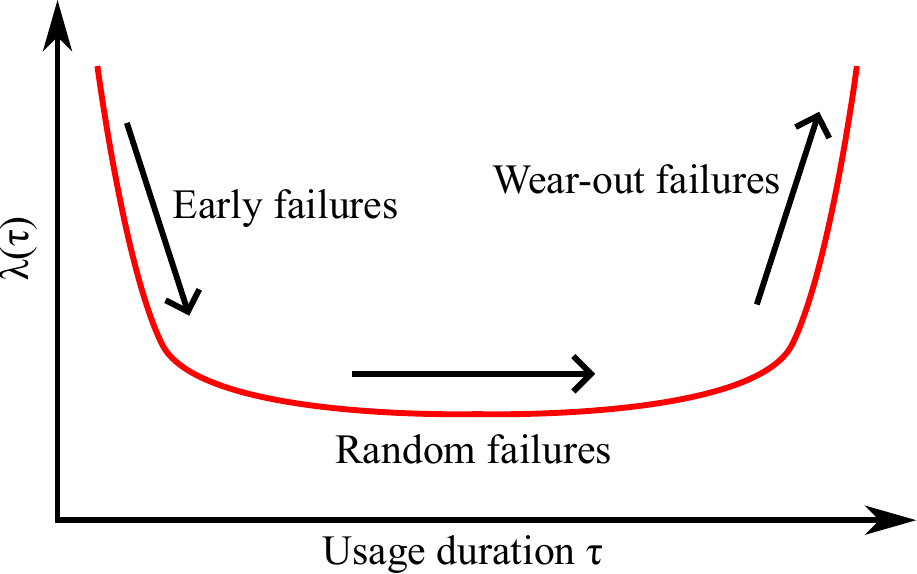}
\caption{Schematic figure of a failure rate $\lambda(\tau)$ of semiconductor products. It is known that the $\lambda(\tau)$ exhibits a bathtub-like failure rate as a function of the usage time $\tau$~\cite{Klutke_2003}. Initially, the $\lambda(\tau)$ decreases because the products continue to fail due to defective products with a certain extent, and then becomes constant. The former period is called the early failure period, and the latter is the random failure period. Finally, the $\lambda(\tau)$ rapidly increases as the products deteriorate due to wear. This is called the wear-out period.} 
\label{failure_rate_schmmatic}
\end{figure}

We first consider a failure probability density $f(\tau)$ representing the probability density that a failure will occur between $\tau$ and $\tau+d\tau$. A failure rate $\lambda(\tau)$ is defined in terms of $f(\tau)$ as 
\begin{align}
    \lambda (\tau)=\frac {f(\tau)}{R(\tau)},
    \label{Failure_rate}
\end{align}
where $R(\tau)$ represents survival function defined as
\begin{align}
    R(\tau)= \int^{\infty}_{\tau} f(t) dt =1-F(\tau)
    \label{Survival function}
\end{align}
with $F(\tau)$ being the failure distribution function as   
\begin{align}
    F(\tau)= \int ^{\tau} _0 f(t) dt.
    \label{f_cdf}
\end{align}
The failure rate $\lambda (\tau)$ describes the frequency of breaking down after the usage time $\tau$. Based on $\lambda(\tau)$, the nature of the failures in semiconductor products can properly classified:  Figure~\ref{failure_rate_schmmatic} shows typical behavior of $\lambda(\tau)$, called bathtub-like curve; initially, the $\lambda(\tau)$ decreases because the initial defective products with a certain extent continue to fail. This period is called the early failure period. Then, the $\lambda(\tau)$ becomes constant. This period is called the random failure period. Finally, the $\lambda(\tau)$ rapidly increases as the product deteriorates due to wear. This is wear-out period. 

In Sec.~\ref{PON_and_POFF}, we confirmed that the probability density of the ON duration follows the exponential function of $P(\tau_{{\rm ON}})$ in Eq.~(\ref{expon}) and that of the OFF duration follows the log-normal function $P(\tau_{{\rm OFF}})$ in Eq.~(\ref{lognorm}). Now, these $P(\tau_{{\rm ON}})$ and $P(\tau_{{\rm OFF}})$ correspond to $f(\tau)$ in Eq.~(\ref{Failure_rate}): Thus, the $\lambda(\tau_{{\rm ON}})=P(\tau_{{\rm ON}})\big/R(\tau_{{\rm ON}})$ describes the failure rate where the ON state breaks to the OFF state after duration $\tau_{{\rm ON}}$. Similarly, the $\lambda(\tau_{{\rm OFF}})$ represents the failure rate where the OFF state changes to the ON state after duration $\tau_{{\rm OFF}}$. By comparing the $\lambda(\tau_{{\rm ON}})$ and $\lambda(\tau_{{\rm OFF}})$ with the bathtub-like curve in Fig.~\ref{failure_rate_schmmatic}, we can discuss the nature of the transition processes of the ON$\to$OFF and OFF$\to$ON. 

\begin{figure}[htpb]
\centering
\includegraphics[width=\linewidth]{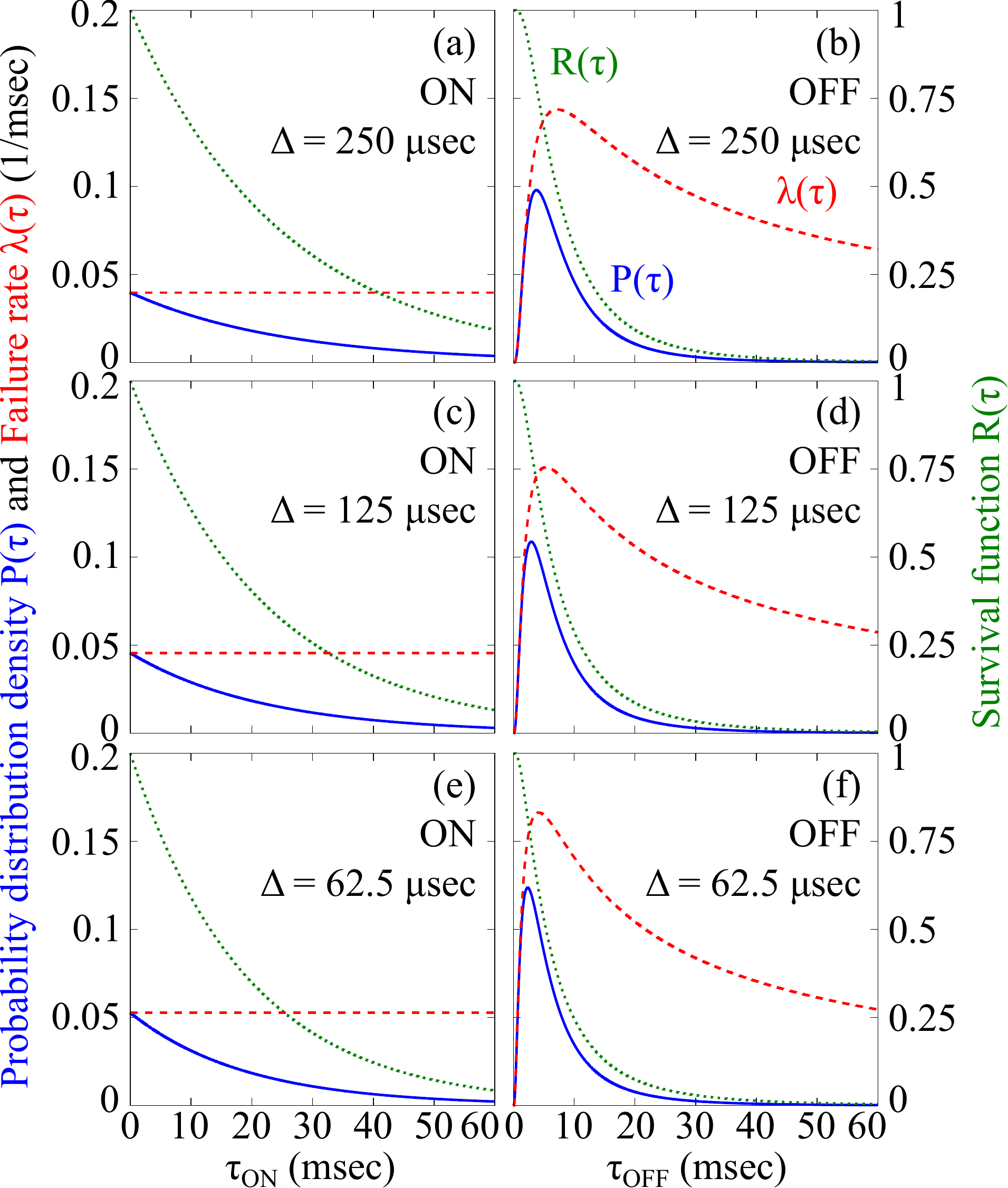}
\caption{Our calculated failure rate $\lambda(\tau)$ in Eq.~(\ref{Failure_rate}) as a function of duration $\tau_{{\rm ON}}$ and $\tau_{{\rm OFF}}$, denoted by red-dashed curves. Survival function $R(\tau)$ in Eq.~(\ref{Survival function}) is described with green-dotted curves. Blue-solid curves are the probability density function $P(\tau)$ displayed in Fig.~\ref{experimental_data_histogram}. The upper (a) and (b), middle (c) and (d), and lower (e) and (f) panels are the results for the time series with $\Delta$ = 250 $\mu$sec, $\Delta$ = 125 $\mu$sec, and $\Delta$ = 62.5 $\mu$sec, respectively. Also, left (a), (c), and (e) panels are results for the ON duration, while right (b), (d), and (f) panels correspond to the results for the OFF duration.} 
\label{pr_sf_fr}
\end{figure}

Figure~\ref{pr_sf_fr} shows our calculated failure rate $\lambda(\tau)$ in Eq.~(\ref{Failure_rate}), denoted by red-dashed curve. Survival function $R(\tau)$ in Eq.~(\ref{Survival function}) is also plotted with green-dotted curve, and blue-solid curve is the model probability density $P(\tau)$ given in Fig.~\ref{experimental_data_histogram}. The upper (a) and (b), middle (c) and (d), the lower (e) and (f) panels are the results for the time series with $\Delta$ = 250 $\mu$sec, $\Delta$ = 125 $\mu$sec, and $\Delta$ = 62.5 $\mu$sec, respectively. Also, the left (a), (c), and (e) panels are results for the ON$\to$OFF process, while the right (b), (d), and (f) panels correspond to the results for the OFF$\to$ON process. We see that the $\lambda(\tau_{{\rm ON}})$ is constant, which is a consequence of the exponential decay of $P(\tau_{{\rm ON}})$. The behavior of $\lambda(\tau_{{\rm ON}})$ is the same as the random failure period in the bathtub-like curve of Fig.~\ref{failure_rate_schmmatic}, and then the ON$\to$OFF process would be recognized as an accidental event; the light-emitting ON state switches to the not-emitting OFF state accidentally.

On the other hand, the OFF$\to$ON process would not be so simple; the $\lambda(\tau_{{\rm OFF}})$ first increases, reaches a maximum, then decreases. The initial increasing trend of the $\lambda(\tau_{{\rm OFF}})$ seems to correspond to the wear-out period in Fig.~\ref{failure_rate_schmmatic}. However, after that, it decreases, which is not observed in the bathtub curve of Fig.~\ref{failure_rate_schmmatic}. In the case of practical semiconductor products, once the wear-out failures begin, all the products would break down within the wear-out period. Therefore, this decreasing trend may not be observed in the  semiconductor-product case. In the present case, the situation might be a little different; at $\tau_{{\rm OFF}}=0$, $P(\tau_{{\rm OFF}})=0$ and the OFF state does not react in the very beginning. Then, the OFF state begins to break down. Around the inflection point of $R(\tau_{{\rm OFF}})$, $\lambda(\tau_{{\rm OFF}})$ takes a peak, and then decreases. The peak position is at around 5 msec, and most of the OFF states return to the ON states in about 30 msec. In any case, when it comes to wear-out failure, various factors, such as the details of the molecular structure and the environment, can be relevant to the failure. 

\section{Summary}\label{sec_summary} 
In this study, we have analyzed the blinking phenomenon of fluorescently labeled DNA using HMM. HMM is an effective method for removing noise in the time series data. The time series after removing noise allows for quantitative determination of the ON (light-emitting) and OFF (not-emitting) states. The length of the fluorescence trajectory is short in the present system, about 1 sec, and the ON and OFF states switches drastically in the trajectory, but HMM works very stably. We have performed analyses for the fluorescence trajectories with different time bins and carefully discussed the time bin dependence of the results. We have found that the probability density of the ON duration is represented by an exponential function, and the OFF-duration probability density is expressed by a log-normal function, which have been verified in terms of the KS test. We have found that the $p$-value of the exponential distribution is 0.2-0.6, and that of the log-normal distribution is $\sim$0.3, and these are larger than the significant level of 0.05. Based on the determined distribution functions, we have performed an analysis based on a failure rate used in the reliability engineering of semiconductor products. We have found that the ON$\to$OFF process is based on a random failure process, and the OFF$\to$ON process in the early stage is similar to a wear-out failure process. 

If the probability density of the OFF duration $\tau_{{\rm OFF}}$ follows the log-normal distribution, the exponent $\alpha_{{\rm OFF}}$ of $\tau_{{\rm OFF}}$ follows the normal distribution (Appendix~\ref{app:normal}). The mean and standard deviation of $\alpha_{{\rm OFF}}$ describe something reaction details of the OFF$\to$ON process; in view of the Arrhenius plot, the mean and standard deviation might be related to the activation energy and its fluctuation with describing the degree of the inhomogeneity effect. More quantitative discussions would be desirable.

In the present analysis for the fluorescence trajectory, we assumed that the noise follows Gaussian distribution (Sec.~\ref{Hidden Markov model}). Although the noise in the time series data often follows Gaussian, the nature of the noise can vary depending on the system and environment. For single-molecule measurements, especially, the nature of the noise is likely to reflect the environmental effects, and machine learning techniques that can handle quantitative nature of the noise, including HMM,  will be important.

\section{Acknowledgments}\label{acknowledgments}
This research was supported by JSPS KAKENHI Grant Numbers JP22H01183, JP23H01126, and JP21H02059, the establishment of university fellowships towards the creation of science technology innovation, Grant Number JPMJFS2133. K.K. and A.M. acknowledge support from the Five-Star Alliance in NJRC Mater. \& Dev.

\appendix
\section{Probability density for logarithm value of the OFF duration}
\label{app:normal}
Here, we discuss the OFF$\to$ON process in terms of the logarithm transform of $\tau$ as  
\begin{eqnarray}
    \alpha=\ln\tau,
    \label{alp}
\end{eqnarray}
where $\alpha$ represents an exponent when the $\tau$ is expressed as an exponential function, and it can be seen as describing the right-hand side of Eq.~(\ref{lntauoff}). The log-normal distribution of $\tau$ in Eq.~(\ref{lognorm}) is rewritten to the normal distribution of $\alpha$ as 
\begin{eqnarray}
    P(\alpha)=\frac {1}{\sqrt{2\pi\sigma^2}} 
    \exp\biggl(-\frac{(\alpha -\mu)^2}{2\sigma^2 }\biggr), 
    \label{normal}
\end{eqnarray}
where we use a transform of $P(\tau)d\tau=P(\alpha)d\alpha$. Thus, the log-normal distribution of $\tau$ is mathematically equivalent to the normal distribution of $\alpha$. It is noted that parameters $\mu$ and $\sigma$ of the above normal distribution are the same as those of the log-normal distribution in Eq.~(\ref{lognorm}). 

\begin{figure}[h!]
\centering
\includegraphics[width=0.7\linewidth]{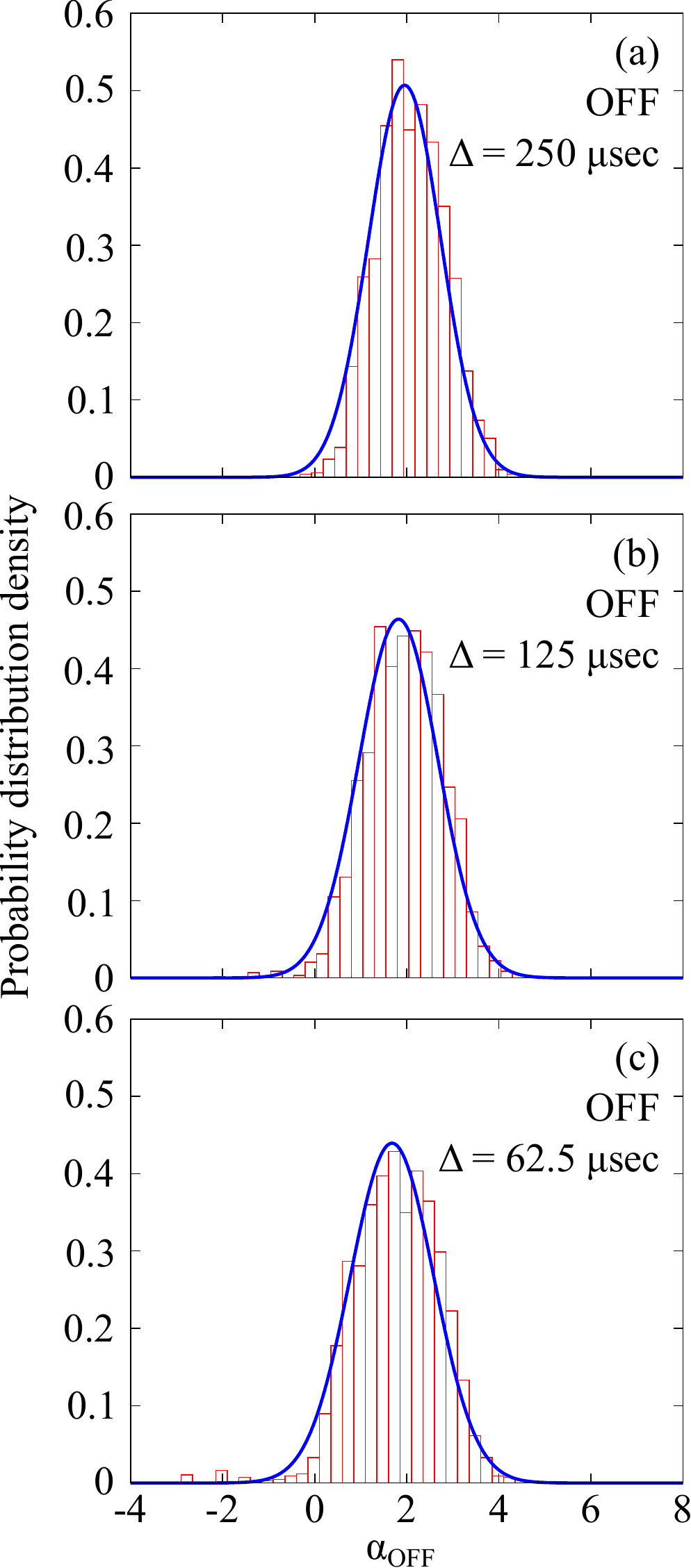}
\caption{Calculated probability density for $\alpha_{{\rm OFF}}=\ln\tau_{{\rm OFF}}$ in Eq.~(\ref{alp}), denoted by red-bars. Blue-solid curves represent normal distributions $P(\alpha_{{\rm OFF}})$ in Eq.~(\ref{normal}), and model parameters and fitting accuracy are summarized in Table~\ref{table_ON_OFF}. The panels (a),  (b), and (c) panels correspond to the results for $\Delta$ = 250 $\mu$sec, 125 $\mu$sec, 62.5 $\mu$sec, respectively.} 
\label{Fig_normal}
\end{figure}

Figure~\ref{Fig_normal} is our calculated probability density for $\alpha$, described by red bars. The grid spacing of $\alpha$ is 0.25, and the total number of the $\alpha$ grids is 21 with the $\Delta$=250-$\mu$sec case, 26 with the $\Delta$=125-$\mu$sec case, and 29 with the $\Delta$=62.5-$\mu$sec case. The blue-solid curves describe the fitted normal distribution functions. Fitting accuracy is summarized in Table~\ref{table_ON_OFF}, where the accuracy is evaluated for the data points of $p(\alpha_k) \ne 0$. 
Figures~\ref{Fig_normal} (a), (b), and (c) correspond to the results for $\Delta$ = 250 $\mu$sec, 125 $\mu$sec, and 62.5 $\mu$sec, respectively. From the figure, we see that the fitting is reasonable, but as the $\Delta$ decreases, the probability around the small $\alpha$ appears. This is related to the short-lived OFF states, and it may be due to quantitative uncertainty of the OFF duration evaluated with the HMM simulation. We note that the KS-$p$ value of the normal distribution $P(\alpha_{{\rm OFF}})$ is the same as that of the log-normal distribution $P(\tau_{{\rm OFF}})$.

%

\end{document}